\documentclass{IEEEoj-data}
\usepackage{cite}
\usepackage{amsmath,amssymb,amsfonts}
\usepackage{algorithmic}
\usepackage{graphicx,color}
\usepackage{textcomp}
\usepackage{morewrites}
\newcounter{auxfile}
\usepackage{url}
\usepackage{float}
\usepackage[colorlinks = true,
            linkcolor = blue,
            urlcolor  = blue,
            citecolor = blue,
            anchorcolor = blue]{hyperref}

\usepackage[most]{tcolorbox}

\def\BibTeX{{\rm B\kern-.05em{\sc i\kern-.025em b}\kern-.08em
    T\kern-.1667em\lower.7ex\hbox{E}\kern-.125emX}}
\AtBeginDocument{\definecolor{ojcolor}{cmyk}{0.93,0.59,0.15,0.02}}

\begin{document}
\receiveddate{00 April, 2024}
\reviseddate{11 April, 2024}
\accepteddate{00 April, 2024}
\publisheddate{00 May, 2024}
\currentdate{7 June, 2024}
\doiinfo{DD.2024.0607000}

\title{{\color{black}{\texttt{BlueTempNet}:}} \textit{A Temporal Multi-network Dataset of Social Interactions in Bluesky Social}}

\author{Ujun Jeong\authorrefmark{1},  Bohan Jiang\authorrefmark{1}, Zhen Tan\authorrefmark{1}, H. Russell Bernard\authorrefmark{2}, Huan Liu\authorrefmark{1}}
\affil{School of Computing and Augmented Intelligence, Arizona State University, Tempe, 85281 USA}
\affil{Institute for Social Science Research, Arizona State University, Tempe, 85281 USA}
\corresp{\{ujeong1, bjiang14, ztan36, asuruss, huanliu\}@asu.edu}
\markboth{IEEE DATA DESCRIPTIONS}{Jeong \textit{et al.}}

\begin{abstract}
Decentralized social media platforms like Bluesky Social (Bluesky) have made it possible to publicly disclose some user behaviors with millisecond-level precision. Embracing Bluesky's principles of open-source and open-data, we present the first collection of the temporal dynamics of user-driven social interactions. BlueTempNet integrates multiple types of networks into a single multi-network, including user-to-user interactions (following and blocking users) and user-to-community interactions (creating and joining communities). Communities are user-formed groups in custom Feeds, where users subscribe to posts aligned with their interests. Following Bluesky's public data policy, we collect existing Bluesky Feeds, including the users who liked and generated these Feeds, and provide tools to gather users' social interactions within a date range. This data-collection strategy captures past user behaviors and supports the future data collection of user behavior.
\\
\\ 
 {\textcolor{ieeedata}{\abstractheadfont\bfseries{IEEE SOCIETY/COUNCIL}}}  IEEE Communications Society (ComSoc)\\  
 \\
 {\textcolor{ieeedata}{\abstractheadfont\bfseries{DATA DOI/PID}}} 10.21227/yrsy-ee91
 \\ 
  
 {\textcolor{ieeedata}{\abstractheadfont\bfseries{DATA TYPE/LOCATION}}} 
 Network Graph; Bluesky Social (Social Media Platform)
 
\end{abstract}


\maketitle
\section*{BACKGROUND}
Understanding the evolution of user behavior through complex interactions in online social networks has been an important topic in computational social science~\cite{blocker2022map, keuschnigg2018analytical, isakov2019structure}, and computer science~\cite{tang2016survey, jin2013understanding, leskovec2010predicting}. However, researchers face two challenges in studying user behavior on social media platforms: (1) platforms typically design recommendations to target users, making it difficult to isolate their effects on user behavior; and (2) platforms often limit access to detailed timing and comprehensive records of how users interact. Table~\ref{tab:data_comparison} categorizes and compares how eight social media platforms manage key aspects of users' social interactions, including \textit{following} (establishing one-way connections and receiving updates from connected users), \textit{blocking} (preventing visibility and engagements between users), \textit{creating} (generating a new community), and \textit{joining} (subscribing to a community's suggested content).
\begin{table}
\caption{Comparing accessible social interactions of users on social media platforms via APIs. The symbols used are: $\times$ for complete exclusion, \(\circ\) for partial access without timestamps, and \(\bullet\) for perfect coverage.}
\label{tab:data_comparison}
\centering
\begin{tabular}{|p{0.3\linewidth}|>{\centering\arraybackslash}m{0.11\linewidth}|>{\centering\arraybackslash}m{0.11\linewidth}|>{\centering\arraybackslash}m{0.11\linewidth}|>{\centering\arraybackslash}m{0.11\linewidth}|}
\hline
\textbf{Platforms} & \multicolumn{2}{c|}{\textbf{User-to-User}} & \multicolumn{2}{c|}{\textbf{User-to-Community}} \\
\cline{2-5}
 & \textbf{Follow} & \textbf{Block} & \textbf{Create} & \textbf{Join} \\
\hline
Twitter/X~\cite{qi2023sentiment, kumar2014twitter}   & \(\circ\) & $\times$ & $\times$ & $\times$ \\
\hline
Facebook~\cite{leskovec2012learning}  & \(\circ\) & $\times$ & $\times$ &  $\times$\\
\hline
Instagram~\cite{yang2021research} &  \(\circ\) & $\times$ & $\times$ &  $\times$\\
\hline
Threads~\cite{jeong2024user} & $\times$ & $\times$ & $\times$ & $\times$ \\
\hline
LinkedIn~\cite{davis2020networking}  & $\times$ & $\times$ & $\times$ &  $\times$\\
\hline
Reddit~\cite{baumgartner2020pushshift}    & $\times$ & $\times$ & \(\circ\) & \(\circ\) \\
\hline
TikTok~\cite{corso2024we}    & $\times$ & $\times$ & $\times$ & $\times$ \\
\hline
Mastodon~\cite{la2021understanding, zignani2018follow}   & \(\circ\) & $\times$ & \(\bullet\)  & \(\bullet\) \\
\hline
\textbf{Bluesky}~\cite{balduf2024looking,quelle2024bluesky}   & \(\bullet\) & \(\bullet\) & \(\bullet\) & \(\bullet\)\\
\hline
\end{tabular}
\end{table}

Interest in alternatives to the central control of traditional social media platforms has led to a substantial migration of users to decentralized platforms~\cite{jeong2024user, failla2024m}. Here, we focus on Bluesky\footnote{\url{https://bsky.social/about}}, a social media platform currently boasting over 5 million active users, where data are distributed and managed across multiple providers. Despite its decentralized architecture, Bluesky offers a seamless user experience comparable to that of centralized platforms. This is achieved through interoperability facilitated by its protocol, ATproto~\cite{kleppmann2024bluesky}. Bluesky features a default reverse chronological timeline and allows users to add customized recommendation algorithms created by other users. This user-driven design facilitates the study of user behavior on social media without the potential control exerted by platforms~\cite{ng2024smi}.

\begin{figure}
\centering
\includegraphics[width=0.5\textwidth]{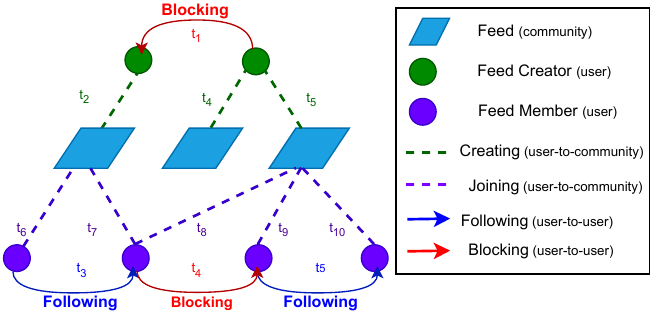}
\caption{Overview of \texttt{BlueTempNet}, a multi-network integrating two categories of users' social interactions such as user-to-user (following, blocking) and user-to-community (creating, joining), all timestamped at $t_i$.}
\label{data_overview}
\end{figure}

Following Bluesky's public data policy\footnote{\url{https://blueskyweb.zendesk.com/hc/en-us/articles/15835264007693-Data-Privacy}} and leveraging the advanced data accessibility and accountability offered through the Personal Data Server (PDS)\footnote{\url{https://en.wikipedia.org/wiki/Personal_data_service}} model, we collected the temporal dynamics of social interactions among over 150,000 public users. Our dataset, \texttt{BlueTempNet}, includes two categories of users' social interactions with millisecond-level timestamps.  Figure~\ref{data_overview} illustrates how these interactions are integrated into a multi-network framework, bridging various types of nodes and edges while allowing multiple directional connections from the same nodes. \texttt{BlueTempNet} offers a unique opportunity for comprehensive analysis, spanning from individual actor networks to the platform ecosystem over time.

The key contributions of our work are:

\begin{itemize}
\item A novel dataset, \texttt{BlueTempNet}, captures the temporal dynamics of users' social interactions, independent of any platform-driven recommendation algorithms.

\item \texttt{BlueTempNet} integrates key behavioral aspects of users' social interactions into a multi-graph with both user-to-user and user-to-community networks.
\item \texttt{BlueTempNet} contains longitudinal data, which is essential for studying the development and sustainability of Bluesky since the introduction of custom Feeds.
\end{itemize}

\begin{figure}
\centering
\includegraphics[width=0.5\textwidth]{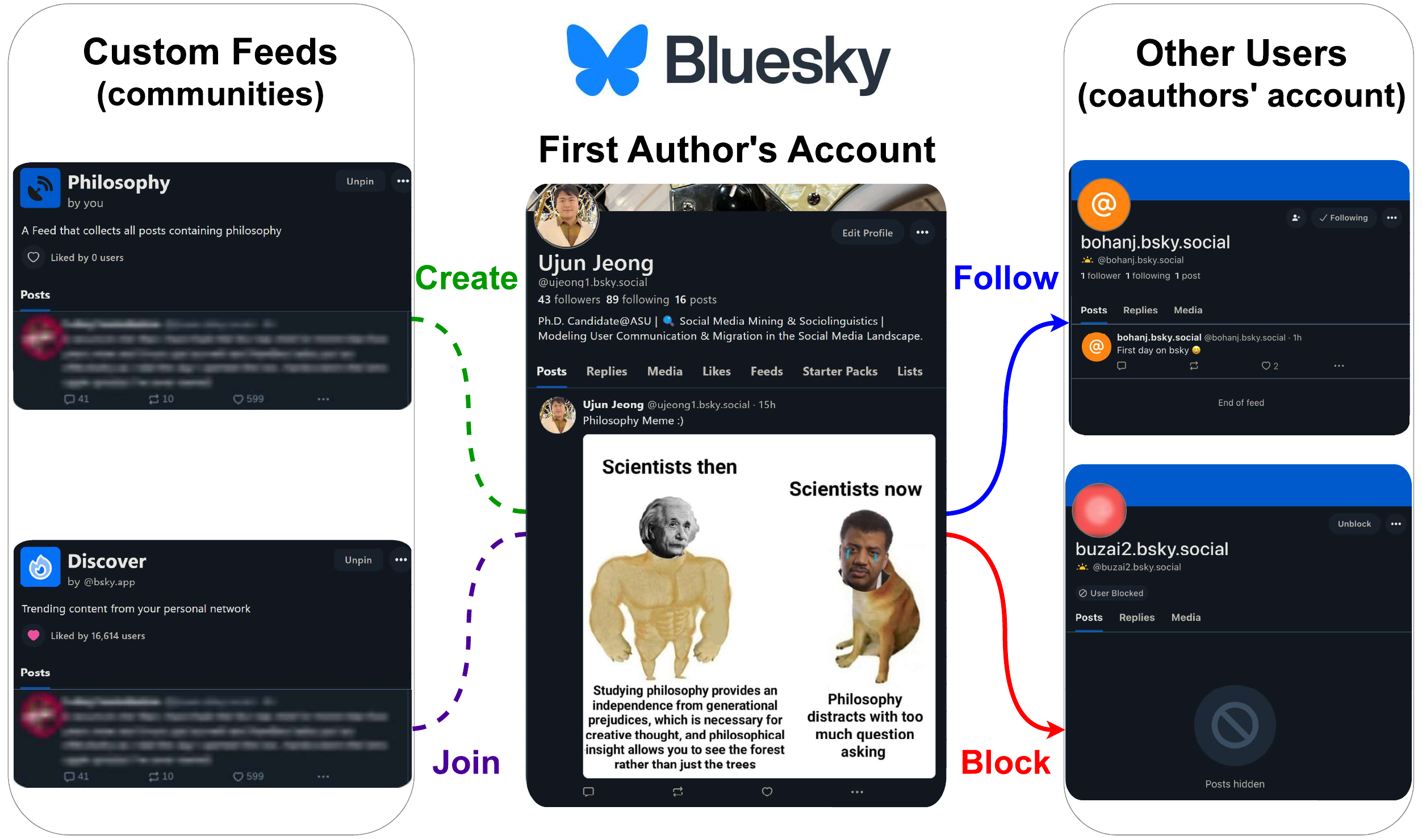}
\caption{An example of a user's interactions on Bluesky. The left panel shows user-to-community interactions, including joining a community (purple dashed line) and a community creation (green dashed line). The right panel illustrates user-to-user interactions, the change in profiles after following (blue arrow) and blocking (red arrow) another user.}\label{bluesky_interfaces}
\end{figure}




\section*{PLATFORM INTERFACE}
 Figure~\ref{bluesky_interfaces} illustrates social interactions using the corresponding authors' account as an example. On the right side of the figure, the \textit{follow} action represents a user's interest in another user, allowing them to receive updates and posts in their default timeline. The \textit{block} feature enhances privacy and safety by ensuring that users who have blocked each other cannot view or engage with each other's posts, making them invisible to both sides. On the left side of the figure, there is a list of \textit{Feeds}, which are dynamic content streams with customizable algorithms. Users create and join \textit{Feeds}, like \texttt{Philosophy} (Philosophy content) and \texttt{Discover} (Trending content). A user creates each feed through ATproto's \textit{Feed Generator} and joins Feeds of interest by clicking the heart icon on the feed's profile, an action called \textit{Feed Like}.
 
\section*{COLLECTION METHODS AND DESIGN}
We used the Bluesky API provided by ATproto\footnote{\url{https://atproto.com/}} protocol. As shown in Figure~\ref{data_collection}, there are three sequential steps in the data collection process: (1) collecting existing Feeds from Bluesky at the community level; (2) gathering information about users who generated or liked these Feeds at the user-to-community level; and (3) compiling relationships between users by examining if they blocked or followed each other at the user-to-user level.
We ensure data quality and retrieval efficiency by using one of REST API endpoints named as \texttt{public.api.bsky.app/xrpc/com.atproto.repo}\break\texttt{.listRecords}, which is publicly accessible without authentication. The records are managed within Bluesky Social's PDS named \texttt{bsky.social}.

\section*{RECORDS AND STORAGE}
\subsection*{STORAGE PRINCIPLES}

We implemented privacy measures, including field-level encryption to anonymize user IDs and the removal of personally identifiable information such as usernames, display names, and biographies, to protect user privacy. At the same time, we adhered to the FAIR~\cite{fair} principles (Findability, Accessibility, Interoperability, and Reusability) by ensuring that the dataset remains reusable and well-documented with appropriate metadata, making it both privacy-compliant and suitable for future research. A decryption key for anonymized user IDs will be made available to bona fide researchers upon request.

\subsection*{SCOPE OF RECORDS}
The versatile data collection framework accommodates different periods of activity established for the research. This study focuses on the year following the introduction of custom Feeds on Bluesky\footnote{\url{https://bsky.social/about/blog/7-27-2023-custom-feeds}}, from May 11, 2023, to May 11, 2024. Based on the time of our data collection.

\begin{figure}
\centering
\includegraphics[width=0.5\textwidth]{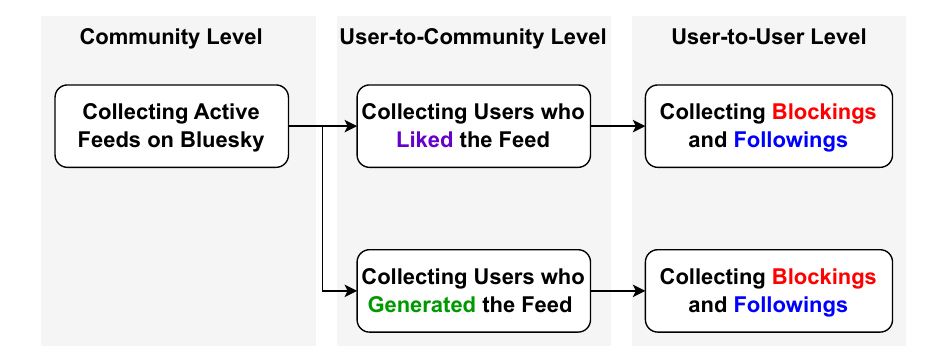}
\caption{An illustration of our data collection pipeline for \texttt{BlueTempNet} is shown in three levels. Each stage of collection captures the timestamp of a user's interactions with millisecond-level precision in UTC format.} 
\label{data_collection}
\end{figure}

\subsection*{USER CATEGORIES}

Though Bluesky does not formally categorize users, we defined two types based on their relationships with Feeds. This simplifies understanding and enhances the hierarchical structure of our proposed multi-graph. These two user types are defined as follows:

\begin{enumerate}
\item \textbf{Feed Creators:} Users who create Feeds using the Feed generator. These users can provide content for others by proposing their own recommendation algorithms.


\item \textbf{Feed Members:} Users who join Feeds by liking them. These users primarily consume the curated content provided by the Feed's recommendation algorithm.
\end{enumerate}

Table~\ref{table:bluesky_stats} provides basic statistics for Bluesky, including the number of users (comprising Feed creators and members), the number of Feeds, and some engagement metrics associated with their profiles. It is important to note that these numbers do not represent our collected dataset. Subsequent section details our specific method on collecting the interactions between Feed creators and members. 

\begin{table}[h!]
\centering
\caption{Overview of total users and Feeds on Bluesky as of May 11, 2024. Profile metrics, such as \#Likes, are aggregated to show overall totals.}\renewcommand{\arraystretch}{1.3}
\setlength{\tabcolsep}{10pt}
\begin{tabular}{|p{0.68\linewidth}|p{0.14\linewidth}|}
\hline
\textbf{Types} & \textbf{Value} \\
\hline
\#Users & 5,659,340 \\
\hline
\#Feeds & 39,968 \\
\hline
\#Likes on Feeds & 322,930 \\
\hline
\#Feed Creators (unique users who have generated at least one Feed) & 17,617 \\
\hline
\#Feed Members (unique users who have liked at least one Feed) & 147,577 \\
\hline

\end{tabular}
\label{table:bluesky_stats}
\end{table}

\subsection*{DIMENSIONS OF RECORDS}
Figure~\ref{data_collection} illustrates the data collection process of \texttt{BlueTempNet}. Based on the user types we categorized, we propose three network dimensions to elicit the hierarchical structure of our dataset and their various interactions.

\begin{enumerate}
\item \textbf{Feed Creator Interaction Network}: This is a signed, directed graph representing interactions between Feed creators. It is defined as $G_{\mathcal{C}} = \{\mathcal{C}, \mathcal{E}^{+}, \mathcal{E}^{-}\}$, where $\mathcal{C}$ denotes the set of Feed creators, $\mathcal{E}^{+}$ represents the set of positive edges (following), and $\mathcal{E}^{-}$ represents the set of negative edges (blocking).
\item \textbf{Feed Member Interaction Network}: This is a signed, directed graph representing interactions between Feed members. It is defined as $G_{\mathcal{M}} = \{\mathcal{M}, \mathcal{E}^{+}, \mathcal{E}^{-}\}$, where $\mathcal{M}$ denotes the set of Feed members, $\mathcal{E}^{+}$ represents the set of positive edges (following), and $\mathcal{E}^{-}$ represents the set of negative edges (blocking).
\item \textbf{Community Interaction Network}: This is affiliation graph that connects communities of users to Feeds. It is defined as $G_{\mathcal{A}} = \{\mathcal{V}, \mathcal{E}\}$, where the set of nodes $\mathcal{V} = \{\mathcal{C}, \mathcal{M}, \mathcal{F}\}$ consists of three distinct types: Feed creators, Feed members, and Feeds. The set of edges $\mathcal{E} = \{\mathcal{G}, \mathcal{L}\}$ consists of two undirected edge types: generation, denoted by $\mathcal{G}$, and likes, denoted by $\mathcal{L}$.

\end{enumerate}

Following and blocking are not mutually exclusive actions.  A user can follow others initially and then block them later. For these cases, we prioritize the blocking interaction by retaining the negative edges (blocking) and removing the positive edges (following). Based on this, we present the statistics for the three dimensions of the dataset: (1) Table~\ref{first_dataset} shows the Feed creator interaction network. The graph contains $1,749$ isolated nodes and there were $485$ edges with both negative and positive signs. (2) Table~\ref{second_dataset} presents the Feed member interaction network. The graph includes $9,145$ isolated nodes and there were $5,502$ edges with both negative and positive signs. (3)  Table~\ref{third_dataset} describes the community interaction network. In the affiliation graph, we identified $8,527$ users who act as both Feed creators and Feed members. These users are represented as two nodes: one as a member node and the other as a creator node.




\begin{figure}
\centering
\includegraphics[width=0.5\textwidth]{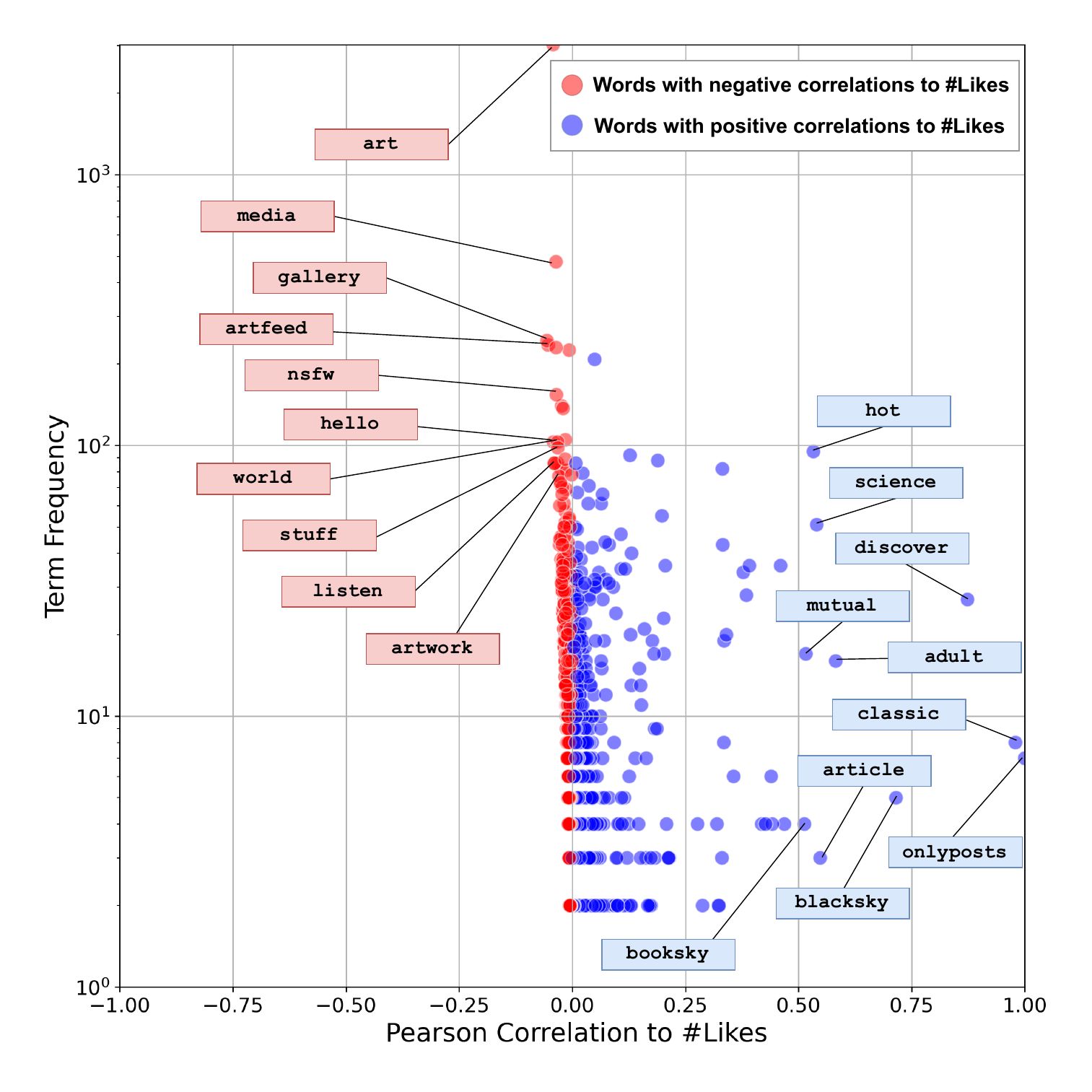}
\caption{Scatter plot displaying the term frequency of terms in Feeds display names (y-axis) and their Pearson correlations with the number of likes received (x-axis). Scatter points are color-coded based on the direction of the correlation. We tagged the points for the top 10 highest and lowest Pearson correlations by displaying the corresponding terms.}
\label{Feedcorrelation}
\end{figure}

\begin{table}
\caption{First dimension -- statistics of the signed directed graph $G_{\mathcal{C}}$, showing block and follow among the Feed creators identified in Table~\ref{table:bluesky_stats}.}
\centering
\renewcommand{\arraystretch}{1.3}
\setlength{\tabcolsep}{10pt}
\begin{tabular}{|p{0.68\linewidth}|p{0.14\linewidth}|}
\hline
\textbf{Types} & \textbf{Count} \\
\hline
Node (Feed Creators accessible through API) & 17,146 \\
\hline
Negative Edge (Blocking among Feed Creators) & 24,362 \\
\hline
Positive Edge (Following among Feed Creators) & 273,696 \\
\hline
\end{tabular}
\label{first_dataset}
\end{table}

\begin{table}
\caption{Second dimension -- statistics of the signed directed graph $G_{\mathcal{M}}$, showing block and follow among the Feed members identified in Table~\ref{table:bluesky_stats}.}
\centering
\renewcommand{\arraystretch}{1.3}
\setlength{\tabcolsep}{10pt}
\begin{tabular}{|p{0.68\linewidth}|p{0.14\linewidth}|}
\hline
\textbf{Types} & \textbf{Count} \\
\hline
Node (Feed Members accessible through API) &  134,946\\
\hline
Negative Edge (Blocking among Feed Members) & 435,700
\\
\hline
Positive Edge (Following among Feed Members) & 4,871,132 \\
\hline
\end{tabular}
\label{second_dataset}
\end{table}

\begin{table}
\caption{Third dimension -- statistics of the affiliation graph $G_{\mathcal{A}}$, showing the relations across Feeds, creators, and members in Tables~\ref{first_dataset} and~\ref{second_dataset}.
}
\centering
\renewcommand{\arraystretch}{1.3}
\setlength{\tabcolsep}{10pt}
\begin{tabular}{|p{0.68\linewidth}|p{0.14\linewidth}|}
\hline
\textbf{Types} & \textbf{Count} \\
\hline
Creator Node (Feed Creators) & 17,536 \\
\hline
Member Node (Feed Members) & 136,724\\
\hline
Feed Node (Feeds accessible through API)  & 39,235 \\
\hline
Join Edge (Users Liking Feeds) & 297,828 \\
\hline
Create Edge (Users Generating Feeds)  & 39,235 \\


\hline
\end{tabular}
\label{third_dataset}
\end{table}

\section*{INSIGHTS AND NOTES}
\subsection*{FEED ECONOMIC DYNAMICS}

Figure~\ref{Feedcorrelation} illustrates the most common terms used in Feed display names and their correlation with popularity, measured by likes. This analysis used term frequency (TF) vectors, defined as \( \text{TF}(t, d) = \frac{\text{count}(t, d)}{\sum_{t' \in d} \text{count}(t', d)} \), where \( t \) is the term and \( d \) is the display name. Pearson correlation coefficients were calculated between the occurrence of these terms and the number of likes received by Feeds containing them. Interestingly, terms such as \texttt{art}, \texttt{media}, and \texttt{gallery}, though commonly used in display names, showed little to no correlation with popularity, with most correlations falling below $-0.005$, suggesting a negligible association with Feed popularity. In contrast, terms like \texttt{likes}, \texttt{discover}, \texttt{classic}, and \texttt{hot} exhibited strong positive correlations with Feed popularity. In addition, we found that terms associated with specific communities, such as \texttt{blacksky} (Black community), \texttt{science} (academic communication), \texttt{booksky} (book club), and \texttt{adult} content, were also positively correlated with higher numbers of likes.


\begin{tcolorbox}[colback=black!5!white,colframe=black!75!black,title=Summary (Feed Economic Dynamics),rounded corners]
There is a contrast between the niche Feeds users create and those that gain widespread popularity. On Bluesky, while art and media Feeds are common, those centered on content discovery, trending topics, and certain communities often receive the most likes.
\end{tcolorbox}



\subsection*{TEMPORAL INTERACTION DYNAMICS}
\begin{figure}
\centering
\includegraphics[width=0.457\textwidth]{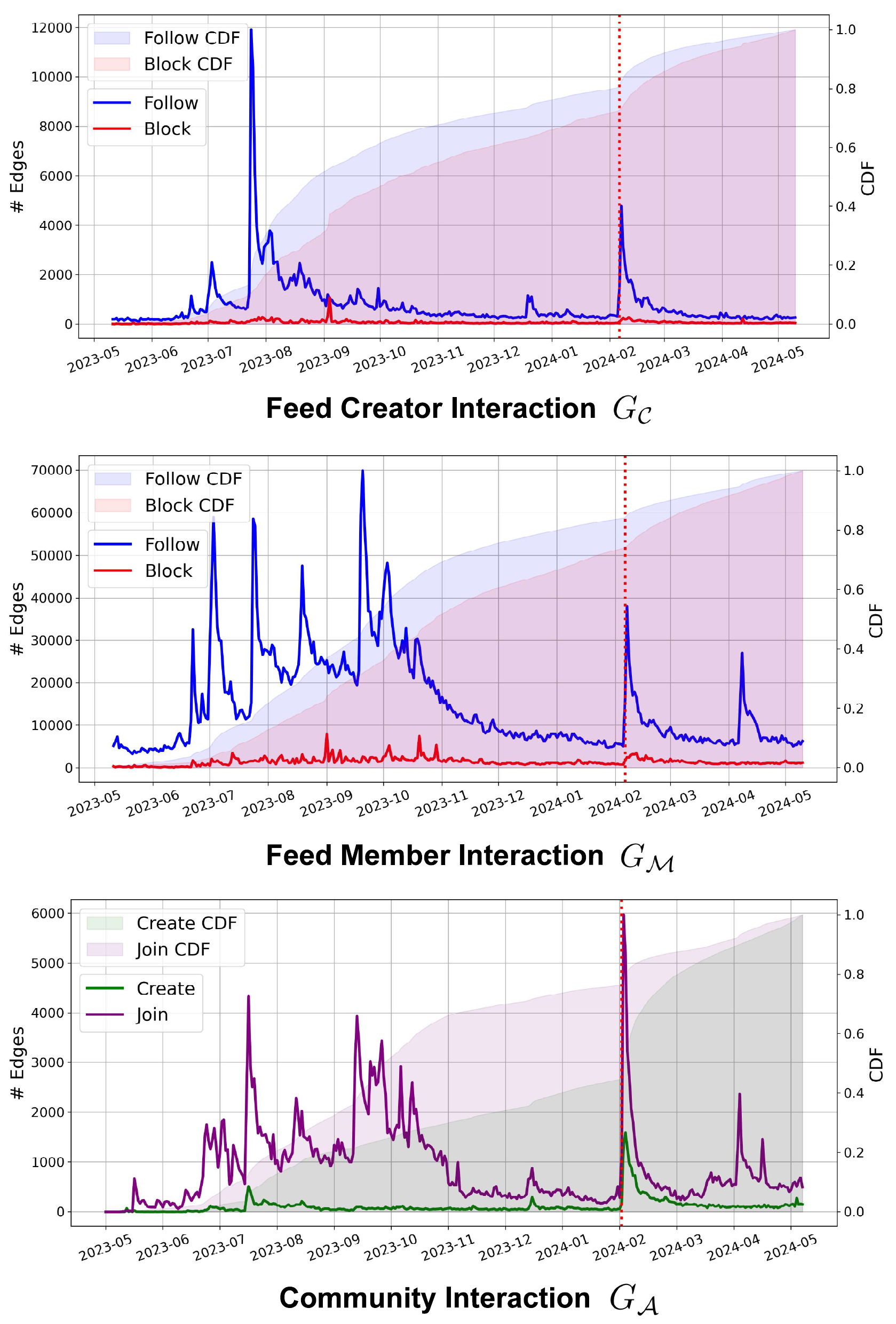}
\caption{Trends in the increase of new edges across three networks: \( G_\mathcal{C} \), \( G_\mathcal{M} \), and \( G_\mathcal{A} \). The lines represent the number of new edges added on each date, while the shaded area indicates the cumulative distribution function (CDF) for each interaction type. The red dotted line marks Bluesky's transition to an invitation-free platform.}
\label{temporal_edge_analysis}
\end{figure}

Figure~\ref{temporal_edge_analysis} highlights the trends in three dimensions of social interactions in \texttt{BlueTempNet} after the platform's shift to an invitation-free model on February 6, 2024. From February 6 to the conclusion of our analysis, the cumulative distribution shows that the number of follows increased by $19.2\%$ and blocks by $27.3\%$ in the first dimension. In the second dimension, follows rose by $15.8\%$ and blocks by $25.9\%$. The third dimension indicates a remarkable $44\%$ increase in the creation of Feeds, along with $34.12\%$ rise in joining Feeds through likes. This analysis demonstrates that blocking users and creating Feeds became more pronounced than their counterpart during the the platform's public launch.

\begin{figure}
\centering
\includegraphics[width=0.46\textwidth]{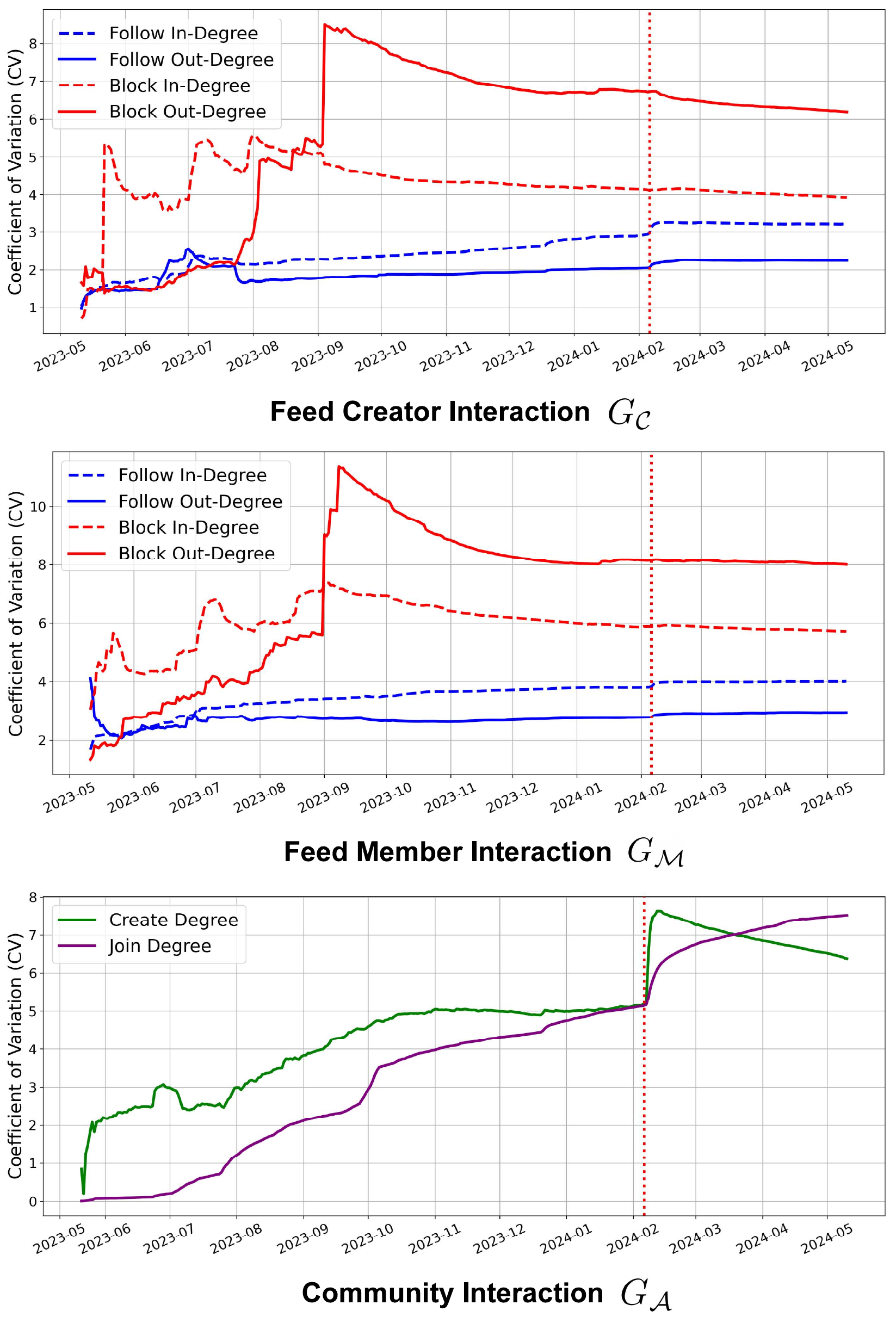}
\caption{Trends of the coefficient of variation (CV) based on degrees of nodes across three networks: \( G_\mathcal{C} \), \( G_\mathcal{M} \), and \( G_\mathcal{A} \). Solid and dashed lines indicate in-degree and out-degree, respectively. The red dotted line marks Bluesky's transition to an invitation-free platform.}
\label{temporal_degree_distribution}
\end{figure}

Figure~\ref{temporal_degree_distribution} shows how the distribution of social interactions evolves over time, based on the coefficient of variation (CV) to measure distributional stability. The CV, defined as the ratio of the standard deviation to the mean, increases steadily across all interactions, with a noticeable rise after October 2023. A significant shift follows Bluesky's transition to open beta in February 2024. After the public launch, follow and block interactions remain relatively stable, while create and join interactions show increased variability before stabilizing. This contrast reveals that user-to-user interactions are more resilient than user-to-community interactions during periods of significant user growth.


\begin{tcolorbox}[colback=black!5!white,colframe=black!75!black,title=Summary (Temporal Interaction Dynamics)]
After Bluesky's public launch, social interactions among users surged across all dimensions. Blocks increased at a faster rate than follows, and the creation of Feeds outpaced the rate of users joining Feeds. The coefficient of variation indicated stable growth patterns in follow and block interactions, while create and join interactions showed greater variability. This underscores the distinct characteristics of user behaviors during the platform's grand opening.

\end{tcolorbox}
 

\subsection*{NETWORK ANALYSIS}
We base our network analysis on the static graph from the last data collection on May 11, 2024. For simplicity, we focus on the largest component of each graph, encompassing $89\%$ of nodes in $G_{\mathcal{C}}$, $92\%$ of nodes in $G_{\mathcal{M}}$, and $86\%$ of nodes in $G_{\mathcal{A}}$. To comprehensively view each network's topology, we carefully selected applicable measures for the signed graphs~\cite{cartwright1956structural} and measures for the affiliation graph~\cite{neal2014backbone}.



\subsubsection*{USER-TO-USER (SIGNED GRAPH)}
Table~\ref{tab:signed_network_measures} summarizes metrics for the signed graphs $G_{\mathcal{C}}$ and $G_{\mathcal{M}}$, highlighting five key aspects: (1) Positive assortativity is slightly negative in both graphs, indicating high-degree users tend to follow lower-degree users. (2) Negative degree assortativity is higher for $G_{\mathcal{C}}$ compared to $G_{\mathcal{M}}$, showing distinct blocking patterns between creators and members. (3) Positive reciprocity is higher than negative reciprocity in both graphs, suggesting stronger mutual following than blocking. (4) Both graphs exhibit similar positive clustering coefficients but low negative clustering coefficients, indicating that followers' neighbors are more likely to be connected than blockers' neighbors. (5) $G_{\mathcal{M}}$ shows a slightly higher proportion of balanced triangles than $G_{\mathcal{C}}$, indicating greater prevalence of social balance among Feed members~\cite{facchetti2011computing, cartwright1956structural}.

\begin{table*}
\caption{Global-level metrics of User-to-User networks are provided with values rounded to two decimal places. In each signed graph, path lengths are determined solely using positive edges, as blocking edges prevent visibility and interaction between users, thereby there is no flow of information.}
\label{tab:signed_network_measures}
\centering
\resizebox{\linewidth}{!}{
\begin{tabular}{|m{4cm}|m{8cm}|m{1.5cm}|m{1.5cm}|}
\hline
\textbf{Metric} & \textbf{Description} & \textbf{$G_{\mathcal{C}}$} & \textbf{$G_{\mathcal{M}}$} \\
\hline 
Diameter & The greatest distance between any pair of nodes & 12 & 13 \\ 
\hline
Average Path Length &  The average shortest path between all pairs of nodes &  3.26 & 3.29\\ 
\hline
Degree Density & Proportion of actual edges compared to possible edges & 1.89 $e^{-3}$& 5.10 $e^{-4}$ \\ 
\hline
Positive Degree Assortativity & Bias for nodes to follow neighbors with similar following degrees & -0.08 & -0.06 \\ 
\hline
Negative Degree Assortativity & Bias for nodes to block neighbors with similar blocking degrees & -0.21 & -0.16 \\ 
\hline
Positive Reciprocity & Likelihood of nodes to be mutually linked by following edges & 0.45 & 0.51\\ 
\hline
Negative Reciprocity & Likelihood of nodes to be mutually linked by blocking edges & 0.07 & 0.06 \\ 
\hline
Positive Clustering Coefficient & Average of two adjacent nodes of a follower also follow each other & 0.18 & 0.20\\ 
\hline
Negative Clustering Coefficient &  Average of two adjacent nodes of a blocker also block each other& 0.04 & 0.04\\ 
\hline
Structural Balance of Triangles & Ratio of socially balanced triangles based on undirected edges& 0.47 & 0.52\\ 
\hline
\end{tabular}}
\end{table*}

\begin{table*}
\caption{Global-level metrics for User-to-Community networks where edges are undirected. Values for each metric are rounded to two decimal places.}
\label{tab:community_measures}
\centering
\resizebox{\linewidth}{!}{
\begin{tabular}{|m{4cm}|m{8cm}|m{1.5cm}|m{1.5cm}|}
\hline
\textbf{Metric} & \textbf{Description} & $G_\mathcal{A}$ & $P_\mathcal{A}$ \\
\hline
Diameter & The greatest distance between any pair of nodes & 21 & 10 \\ 
\hline
Average Path Length & The average shortest path between all pairs of nodes & 5.30 & 3.25 \\ 
\hline
Degree Density & Proportion of actual edges compared to possible edges & 2.29 $e^{-5}$ &  1.40 $e^{-2}$\\ 
\hline
Degree Assortativity & Nodes to connect with similar degrees of neighborhood & -0.16 & 0.59  \\ 
\hline
Clustering Coefficient & Average of two neighbors of a node are also neighbors to each other & 0.00 & 0.77\\ 
\hline
\end{tabular}}
\end{table*}

\subsubsection*{USER-TO-COMMUNITY (AFFILIATION GRAPH)}
To analyze higher-order relationships between communities, we define a line graph \( P_{\mathcal{A}} = \{\mathcal{F}, \mathcal{E}, \mathcal{W}\} \) by projecting the affiliation graph \( G_{\mathcal{A}} \) as follows:

\begin{equation}
\begin{aligned}
    \mathcal{E} &= \left\{ (f_i, f_j) \mid U(f_i) \cap U(f_j) \neq \emptyset \right\}\\
    w(f_i, f_j) &= \left| U(f_i) \cap U(f_j) \right|, \:\text{with}\: w \in \mathcal{W}
\end{aligned}
\end{equation}

\noindent where \( U(f_i) \) denotes the set of users adjacent to one of Feeds \( f_i \in \mathcal{F} \) from \( G_{\mathcal{A}} \). Two nodes \( f_i \) and \( f_j \) are linked if they share at least one user. The weight of an edge \( w(f_i, f_j) \) is the number of shared users between the Feeds. 

Table~\ref{tab:community_measures} compares the affiliation graph \( G_{\mathcal{A}} \) with its projected line graph \( P_{\mathcal{A}} \), revealing three key differences: (1) the degree density is higher in \( P_{\mathcal{A}} \) than in \( G_{\mathcal{A}} \). (2) \( P_{\mathcal{A}} \) exhibits a short average path length and a high clustering coefficient, which is close to the maximum, indicating the properties of a small-world network~\cite{watts1998collective}. (3) the degree assortativity for \( G_{\mathcal{A}} \) is negative, whereas it is positive in \( P_{\mathcal{A}} \), showing that Feeds with more members tend to connect together.

\begin{figure*}
\centering
\includegraphics[width=1.0\textwidth]{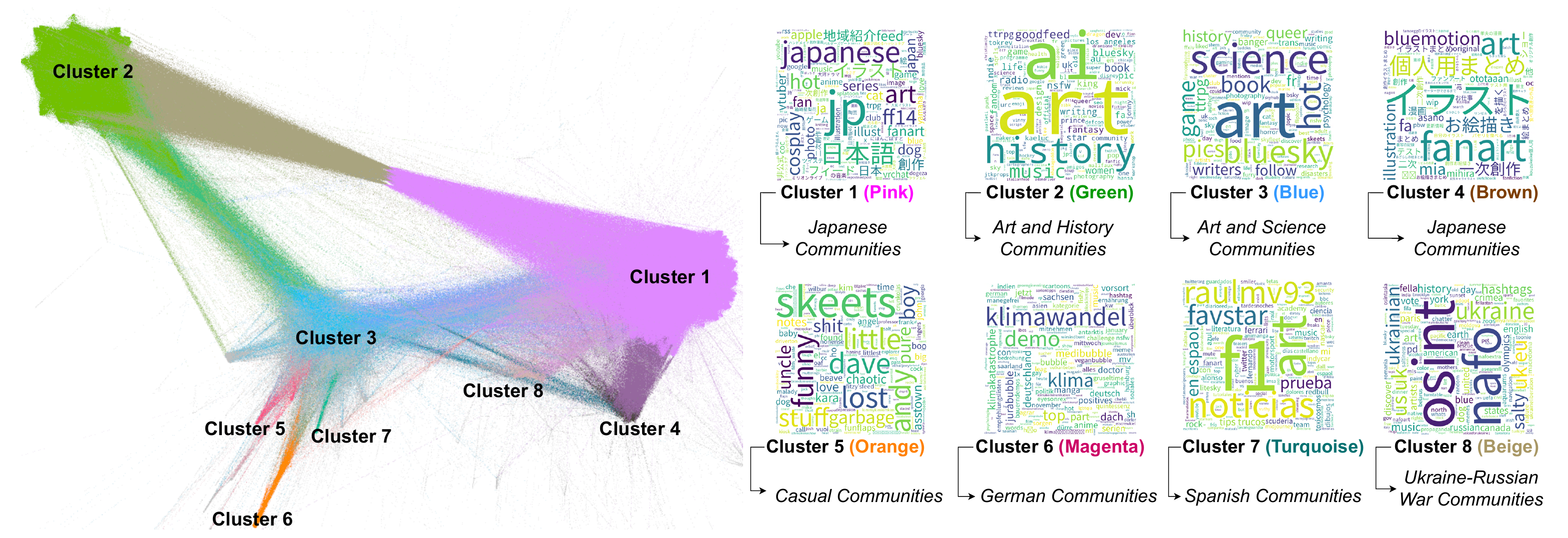}
    \caption{
A line graph of Feeds $P_{\mathcal{A}}$ is visualized using ForceAtlas2. A total of 770 clusters are identified by the Leiden Algorithm. The top 8 clusters are color-coded; others are gray. Each cluster's interests are shown by word clouds, which aggregate the Feeds' self-descriptions in each cluster.}
\label{graph_visualize}
\end{figure*}

Figure~\ref{graph_visualize} presents the community analysis of the projected line graph \( P_{\mathcal{A}} \), where clusters are identified through the Leiden Algorithm~\cite{traag2019louvain} designed for a weighted graph. Each cluster is visualized through word clouds generated from the descriptions in the Feeds, highlighting communities based on different languages and regions, including German, Japanese, Spanish, and English. Furthermore, the clusters features communities for both political and casual topics, such as the Ukraine-Russia war and art-related communities. As discussed during network analysis, this projected line graph exhibits small-world properties. Further exploration could uncover tightly knit community groups, including higher-order analyses to identify strongly connected clusters and examine their characteristics with respect to language, political geography, and relevant social events.
\begin{tcolorbox}[colback=black!5!white,colframe=black!75!black,title=Summary (Network Analysis)]
Both creators and members exhibit negative degree assortativity in their followings, primarily linking to nodes with lower degrees. Additionally, members often connect blockers more effectively, which highlights distinct user roles and higher structural balance. Projecting the affiliation graph revealed small-world properties within Feeds, particularly clustered around communities related to art and languages.
\end{tcolorbox}

\subsection*{DISCUSSION}
Bluesky presents a unique opportunity for researchers to examine user behavior on social media, but it also faces significant challenges to its long-term viability. As the platform is still developing, user engagement may be less consistent than on established platforms like Twitter/X, resulting in unstable usage patterns influenced by updates and external factors. Users tend to focus on niche topics like art, which may limit the generalizability of our findings to other social media platforms. Our analysis specifically targeted networking behaviors, excluding interactions related to posts, such as replying, liking, and reposting.

Furthermore, previous research highlights that external factors, such as interactions across different social media platforms, can affect user retention on Bluesky~\cite{jeong2024user}. A key issue is that Bluesky is often viewed as a complementary platform to Twitter/X rather than a direct competitor. This perception influences user preferences, as many migrating users demonstrate strong loyalty to their original platforms, particularly Twitter/X. Such platform inertia poses a significant barrier to Bluesky's ability to retain and engage users.

In future research, we will explore the evolutionary dynamics of sentiments and rumors on the platform \cite{jeong2022nothing, jeong2022classifying}, focusing on how sentiment evolves as information spreads over time \cite{kramer2014experimental}. Additionally, we will investigate the effects of specific networking behaviors, such as user-to-user blocking \cite{ghasemian2024structure, isakov2019structure}, on the platform's resilience during periods of significant change \cite{jeong2024exploring, jeong2024user}. Our study will evaluate how blocking influences social networking dynamics and platform stability, particularly as new users join. Measuring the global structural balance of social interactions, along with the causal mechanisms behind these networking patterns, can provide valuable insights into how the platform's environment affects user behavior~\cite{facchetti2011computing}. Last, we aim to identify consistent user interaction patterns across a variety of online social networks, including those with and without recommendation systems \cite{avalle2024persistent, leskovec2010predicting}.

\section*{SOURCE CODE AND SCRIPTS}

The scripts and dataset used for crawling Bluesky will be publicly available in the provided DOI repository. The proposed graphs are stored in GEXF\footnote{\url{https://gephi.org/users/supported-graph-formats/}} as follows:
\begin{itemize}
\item \textit{\textbf{graph\_dimension1.gexf}: } Feed member interaction network saved in \texttt{DiGraph} object, where an edge has attributes \texttt{sign} and \texttt{time} and a node is Feed member.
\item \textit{\textbf{graph\_dimension2.gexf}: } Feed creator interaction network saved in \texttt{DiGraph} object, where an edge has attributes \texttt{sign} and \texttt{time}, and a node is Feed creator.
\item \textit{\textbf{graph\_dimension3.gexf}:} Community interaction network saved in a \texttt{Graph} object. Each node has a \texttt{node} attribute that can be one of the following: member, creator, or feed. Each edge has an \texttt{edge} attribute, either join or create, along with a \texttt{time} attribute.

\item \textit{\textbf{multi\_graph.gexf}:} This is a \texttt{MultiGraph} object that integrates the three network dimensions. To facilitate ease of use, all undirected edges in the multigraph have been converted to bidirectional edges.
\end{itemize}

Our data collection scripts are written in Python 3. These scripts utilize the NetworkX library (version $2.6.3$) and specifically the \texttt{networkx.read\_gexf} function\footnote{\url{https://networkx.org/documentation/stable/reference/readwrite/generated/networkx.readwrite.gexf.read_gexf.html}}. This function helps processing different graph objects, including \texttt{DiGraph}, \texttt{Graph}, and \texttt{MultiGraph}. The edge data contains timestamps formatted in the ISO 8601 standard (e.g., \texttt{2023-09-22T09:32:17.974Z}), all recorded in UTC. Our resources are publicly available at our repository\footnote{\url{https://github.com/ujeong1/BlueTempNet-IEEE-DATA-2024}}.


While user IDs and their specific profile information remain anonymous, we provide metadata that includes non-identifiable information about users and details regarding the collected Feeds, as outlined below:
\begin{itemize}
    \item \textit{\textbf{user\_metadata.csv}}
    \begin{enumerate}
        \item Node Index (consistent across all GEXF files)
        \item Anonymized ID (decoded after ID request review)
        \item Number of Followers
        \item Number of Following
        \item Number of Posts
    \end{enumerate}
    \item \textit{\textbf{feed\_metadata.csv}} \begin{enumerate}
        \item Node Index (consistent across all GEXF files)
        \item Feed URI (a unique identifier for profiles specific to the Bluesky Feed)
        \item Display Name of Feed
        \item Description of Feed
        \item Creator of Feed (given as Anonymized ID)
        \item Number of Likes on Feed
    \end{enumerate}

\end{itemize}
\begin{table}[H]
\centering
\caption{Metadata formats in \textit{user\_metadata.csv}. The abbreviation \texttt{b64str} refers to a binary string encoded in Base64 format.}
\label{metadata_user_format}
\begin{tabular}{|c|c|c|}
\hline
\textbf{Metadata} & \textbf{Format} & \textbf{Example} \\
\hline
Node Index & \texttt{integer} & 1\\
\hline
Anonymized ID & \texttt{b64str} & b`gAAAAAB \dots ba66QOmV4'\\
\hline
\#Followers & \texttt{integer} & 23\\
\hline
\#Following & \texttt{integer} & 41\\
\hline
\#Posts & \texttt{integer} & 10\\
\hline
\end{tabular}
\end{table}

\begin{table}[H]
\centering
\caption{Metadata formats in \textit{feed\_metadata.csv}. The abbreviation \texttt{b64str} refers to a binary string encoded in Base64 format.}
\label{metadata_feed_format}
\begin{tabular}{|c|c|c|}
\hline
\textbf{Metadata} & \textbf{Format} & \textbf{Example} \\
\hline
Node Index & \texttt{integer} & 3\\
\hline
Feed URI & \texttt{string} & did:plc:\dots .feed.generator/philosophy \\
\hline
Display Name & \texttt{string} & philosophy\\
\hline
Description & \texttt{string} & Feed for philosophy on Bluesky! \dots\\
\hline
Creator & \texttt{b64str} & b‘gAAAAAB . . . ba6QOmV4’\\
\hline
\#Likes & \texttt{integer} & 10\\
\hline
\end{tabular}
\end{table}

Metadata formats are detailed in Tables~\ref{metadata_user_format} and \ref{metadata_feed_format}. For clarity, examples are abbreviated with ellipses. Each row in the metadata CSV files features a unique node index alongside a user ID or Feed URI. The node index remains consistent across graph dimensions, facilitating searching nodes. A user can appear in multiple rows as both a creator and a member; for instance, the anonymized user ID \texttt{b`gAAAAAB \dots ba66QOmV4'} may correspond to both node index 1 and node index 2. In contrast, each node index uniquely identifies a single feed; for example, node index 3 is exclusively linked to the Feed URI \texttt{did:plc:\dots .feed.generator/philosophy}.


\section*{ACKNOWLEDGMENTS AND INTERESTS}

This work was funded by Office of Naval Research, under Award No. N00014-21-1-4002. Opinions, interpretations, conclusions, and recommendations within this article are solely those of the authors. The article authors have declared no conflicts of interest.


\bibliographystyle{splncs04}
\bibliography{references}

\begin{thebibliography}{10}
\providecommand{\url}[1]{\texttt{#1}}
\providecommand{\urlprefix}{URL }
\providecommand{\doi}[1]{https://doi.org/#1}

\bibitem{avalle2024persistent}
Avalle, M., Di~Marco, N., Etta, G., Sangiorgio, E., Alipour, S., Bonetti, A., Alvisi, L., Scala, A., Baronchelli, A., Cinelli, M., et~al.: Persistent interaction patterns across social media platforms and over time. Nature  (2024)

\bibitem{balduf2024looking}
Balduf, L., Sokoto, S., Ascigil, O., Tyson, G., Scheuermann, B., Korczy{\'n}ski, M., Castro, I., Kr{\'o}l, M.: Looking at the blue skies of bluesky. arXiv preprint arXiv:2408.12449  (2024)

\bibitem{baumgartner2020pushshift}
Baumgartner, J., Zannettou, S., Keegan, B., Squire, M., Blackburn, J.: The pushshift reddit dataset. In: ICWSM (2020)

\bibitem{blocker2022map}
Bl{\"o}cker, C., Nieves, J.C., Rosvall, M.: Map equation centrality: community-aware centrality based on the map equation. Applied Network Science  (2022)

\bibitem{cartwright1956structural}
Cartwright, D., Harary, F.: Structural balance: a generalization of heider's theory. Psychological review  (1956)

\bibitem{corso2024we}
Corso, F., Pierri, F., De~Francisci~Morales, G.: What we can learn from tiktok through its research api. In: WebSci (2024)

\bibitem{davis2020networking}
Davis, J., Wolff, H.G., Forret, M.L., Sullivan, S.E.: Networking via linkedin: An examination of usage and career benefits. Journal of Vocational Behavior  (2020)

\bibitem{facchetti2011computing}
Facchetti, G., Iacono, G., Altafini, C.: Computing global structural balance in large-scale signed social networks. PNAS  (2011)

\bibitem{failla2024m}
Failla, A., Rossetti, G.: " i'm in the bluesky tonight": Insights from a year worth of social data. arXiv preprint arXiv:2404.18984  (2024)

\bibitem{fair}
{FORCE11}: The fair data principles. \url{https://force11.org/info/the-fair-data-principles/} (2020)

\bibitem{ghasemian2024structure}
Ghasemian, A., Christakis, N.A.: The structure and function of antagonistic ties in village social networks. PNAS  (2024)

\bibitem{isakov2019structure}
Isakov, A., Fowler, J.H., Airoldi, E.M., Christakis, N.A.: The structure of negative social ties in rural village networks. Sociological science  (2019)

\bibitem{jeong2022classifying}
Jeong, U., Alghamdi, Z., Ding, K., Cheng, L., Li, B., Liu, H.: Classifying covid-19 related meta ads using discourse representation through a hypergraph. In: SBP-BRiMS. Springer (2022)

\bibitem{jeong2022nothing}
Jeong, U., Ding, K., Cheng, L., Guo, R., Shu, K., Liu, H.: Nothing stands alone: Relational fake news detection with hypergraph neural networks. In: IEEE Big Data. IEEE (2022)

\bibitem{jeong2024user}
Jeong, U., Nirmal, A., Jha, K., Tang, S.X., Bernard, H.R., Liu, H.: User migration across multiple social media platforms. In: SDM. SIAM (2024)

\bibitem{jeong2024exploring}
Jeong, U., Sheth, P., Tahir, A., Alatawi, F., Bernard, H.R., Liu, H.: Exploring platform migration patterns between twitter and mastodon: A user behavior study. In: ICWSM (2024)

\bibitem{jin2013understanding}
Jin, L., Chen, Y., Wang, T., Hui, P., Vasilakos, A.V.: Understanding user behavior in online social networks: A survey. IEEE communications magazine

\bibitem{keuschnigg2018analytical}
Keuschnigg, M., Lovsj{\"o}, N., Hedstr{\"o}m, P.: Analytical sociology and computational social science. Journal of Computational Social Science  (2018)

\bibitem{kleppmann2024bluesky}
Kleppmann, M., Frazee, P., Gold, J., Graber, J., Holmgren, D., Ivy, D., Johnson, J., Newbold, B., Volpert, J.: Bluesky and the at protocol: Usable decentralized social media. arXiv preprint arXiv:2402.03239  (2024)

\bibitem{kramer2014experimental}
Kramer, A.D., Guillory, J.E., Hancock, J.T.: Experimental evidence of massive-scale emotional contagion through social networks. PNAS  (2014)

\bibitem{kumar2014twitter}
Kumar, S., Morstatter, F., Liu, H.: Twitter data analytics. Springer (2014)

\bibitem{la2021understanding}
La~Cava, L., Greco, S., Tagarelli, A.: Understanding the growth of the fediverse through the lens of mastodon. Applied network science  (2021)

\bibitem{leskovec2010predicting}
Leskovec, J., Huttenlocher, D., Kleinberg, J.: Predicting positive and negative links in online social networks. In: WWW (2010)

\bibitem{leskovec2012learning}
Leskovec, J., Mcauley, J.: Learning to discover social circles in ego networks. NeurIPS  (2012)

\bibitem{neal2014backbone}
Neal, Z.: The backbone of bipartite projections: Inferring relationships from co-authorship, co-sponsorship, co-attendance and other co-behaviors. Social Networks  (2014)

\bibitem{ng2024smi}
Ng, L.H.X., Phillips, S.C., Carley, K.M.: Smi-5: Five dimensions of social media interaction for platform (de) centralization. ICWSM workshop  (2024)

\bibitem{qi2023sentiment}
Qi, Y., Shabrina, Z.: Sentiment analysis using twitter data: a comparative application of lexicon-and machine-learning-based approach. Social Network Analysis and Mining  (2023)

\bibitem{quelle2024bluesky}
Quelle, D., Bovet, A.: Bluesky: Network topology, polarisation, and algorithmic curation. arXiv preprint arXiv:2405.17571  (2024)

\bibitem{tang2016survey}
Tang, J., Chang, Y., Aggarwal, C., Liu, H.: A survey of signed network mining in social media. ACM Computing Surveys (CSUR)  (2016)

\bibitem{traag2019louvain}
Traag, V.A., Waltman, L., Van~Eck, N.J.: From louvain to leiden: guaranteeing well-connected communities. Scientific reports  (2019)

\bibitem{watts1998collective}
Watts, D.J., Strogatz, S.H.: Collective dynamics of ‘small-world’networks. nature  (1998)

\bibitem{yang2021research}
Yang, C.: Research in the instagram context: Approaches and methods. The Journal of Social Sciences Research  (2021)

\bibitem{zignani2018follow}
Zignani, M., Gaito, S., Rossi, G.P.: Follow the “mastodon”: Structure and evolution of a decentralized online social network. In: ICWSM (2018)

\end{thebibliography}

\end{document}